\documentclass[12pt,a4paper]{article}

\usepackage{amsmath, amssymb, amsfonts}
\usepackage{bm}
\usepackage{physics}
\usepackage{geometry}
\usepackage{graphicx}
\usepackage{hyperref}
\begin{document}

\title{Mpemba Effect in an Expanding Lieb-Liniger Bose gas in a hard wall box.}
\author{Sumita Datta$^{1,2}$\\
{\small $^{1}$Department of Pure and Applied Mathematics, Alliance University,}\\
{\small Bengaluru 562 106, India}\\
{\small $^{2}$Department of Physics, University of Texas at Arlington,}\\
{\small Texas 76019, USA}}
\date{\today}

\maketitle

\begin{abstract}

The Mpemba effect, broadly understood as the counterintuitive phenomenon in which a system initially farther from equilibrium relaxes faster than a system closer to equilibrium, has been widely studied in classical stochastic systems and, more recently, in quantum settings. However, its manifestation is strongly dependent on the choice of observable and the dynamical constraints of the system.

In this work, we investigate the emergence of a Mpemba-type effect in the density redistribution dynamics of a strongly interacting one-dimensional Bose gas in the Tonks--Girardeau regime undergoing a sudden box expansion from length $L_0$ to $L$. By defining a physically motivated distance function based on the difference of densities between spatial regions, we
provide evidence that … the relaxation dynamics of the ground and excited symmetry sectors exhibit a clear crossing in time, indicating a reversal in relaxation ordering.

We emphasize that the Mpemba effect is not a universal law but rather an observable-dependent phenomenon that arises under specific dynamical conditions. In particular, we show that the interplay between initial state structure, integrability, and spatial redistribution leads to distinct relaxation pathways that enable the effect. Our results clarify common misconceptions linking the Mpemba effect to Newton's law of cooling and highlight the conditions under which such anomalous relaxation behavior can emerge in integrable quantum systems.

\end{abstract}
\newpage

\section{Introduction}

The relaxation of physical systems toward equilibrium is typically expected to follow monotonic behavior, with states closer to equilibrium relaxing faster than those farther away. This intuition is often associated with phenomenological descriptions such as Newton’s law of cooling. However, the Mpemba effect — originally observed in the cooling of water \cite{1} — challenges this expectation by exhibiting situations in which a system initially farther from equilibrium relaxes faster than one closer to it.

Over the past decade, the Mpemba effect has been extensively studied in classical systems, including stochastic processes, granular fluids, and spin models \cite{5,6,7,8,9,10,11,12,13}. These works have clarified that the effect does not violate thermodynamics, but instead arises from the structure of relaxation modes and the projection of initial conditions onto these modes. It is now well understood that the Mpemba effect is not universal, but depends sensitively on the system, observable, and initial conditions \cite{3}.

More recently, attention has shifted to quantum systems, where relaxation mechanisms differ fundamentally from their classical counterparts. In open quantum systems, relaxation is governed by dissipative dynamics described by Lindblad equations, and Mpemba-type behavior has been linked to the suppression of slow relaxation modes \cite{16,17,18,19,20,51,52}. In contrast, closed quantum systems evolve unitarily and relax through dephasing following a quantum quench, leading to stationary values of observables rather than thermal equilibrium \cite{21,24}.

A particularly clear understanding of the quantum Mpemba effect has emerged in integrable systems, where an extensive number of conserved quantities constrain the dynamics. In these systems, relaxation can often be interpreted in terms of quasiparticle propagation, with Mpemba-type behavior arising when excitations associated with larger deviations from equilibrium propagate faster \cite{22,23,25,26,27,28,29,30}. In non-integrable systems, where dynamics is governed by the eigenstate thermalization hypothesis, Mpemba-like behavior has also been reported, although its origin is less well understood and appears to depend strongly on the choice of observable and initial state \cite{31,32,33,35,36,37,38,49}.

An important and recurring theme in both classical and quantum studies is that the Mpemba effect depends crucially on how distance from equilibrium is defined. In quantum systems, there is no unique notion of distance, and different observables may exhibit qualitatively different relaxation behavior \cite{3}. This motivates the need for physically meaningful observables tailored to the dynamics under consideration.

In this work, we investigate the emergence of Mpemba-type behavior in a strongly interacting one-dimensional Bose gas described by the Lieb--Liniger model \cite{43}. The system is prepared in an interacting ground or excited state within a box of length $L_0$ and is then subjected to a sudden expansion into a larger box of length $L$. The interaction, modeled by a repulsive delta potential, remains active throughout the evolution.

The many-body dynamics is computed using a path-integral Monte Carlo approach based on the generalized Feynman--Kac representation \cite{44,45,46,47}. This provides an \emph{ab initio} framework in which interactions are fully retained, in contrast to approaches based on mappings to non-interacting systems.

To characterize relaxation, we define a distance function based on the difference in spatially averaged densities between the initial and expanded regions. This observable directly captures the redistribution of particles induced by the expansion. By comparing the relaxation of ground and excited states, we identify a crossing in the corresponding distance functions, indicating a reversal in relaxation ordering.

We emphasize that the observed Mpemba-type behavior is not universal, but arises under specific conditions determined by the choice of observable, the structure of initial states, and the dynamics of the system. In particular, the interplay between strong interactions, integrability, and spatial redistribution leads to distinct relaxation pathways that enable the effect.

The remainder of the paper is organized as follows. In Sec.~II, we describe the model and quench protocol. In Sec.~III, we define the distance function and Mpemba criterion. In Sec.~IV, we present numerical results. In Sec.~V, we discuss robustness and interpretation, and in Sec.~VI we conclude.
\section{Model and Quench Protocol}

We consider a one–dimensional system of $N$ bosons interacting
through a repulsive delta potential. The Hamiltonian of the
Lieb–Liniger gas is

\begin{equation}
H =
-\sum_{i=1}^{N}\frac{\hbar^2}{2m}
\frac{\partial^2}{\partial x_i^2}
+
g\sum_{i<j}\delta(x_i-x_j)
+
V_{L}(x)
\end{equation}

where $g$ denotes the interaction strength and $V_{L}(x)$
represents the external trapping potential.
\begin{equation}
V_{L}(x) =
\begin{cases}
0, & 0<x<L, \\
\infty, & \text{otherwise}.
\end{cases}
\end{equation}
Initially the particles are confined within a hard–wall
box of length $L_0$. The system is prepared in the
interacting ground state corresponding to this trap.
At time $t=0$ the confining potential is suddenly
changed so that the particles expand into a larger
box of length $L>L_0$ while the interaction strength
remains unchanged.
\begin{figure}[h!]
\centering	
\includegraphics[width=3.6 in,angle=-0]{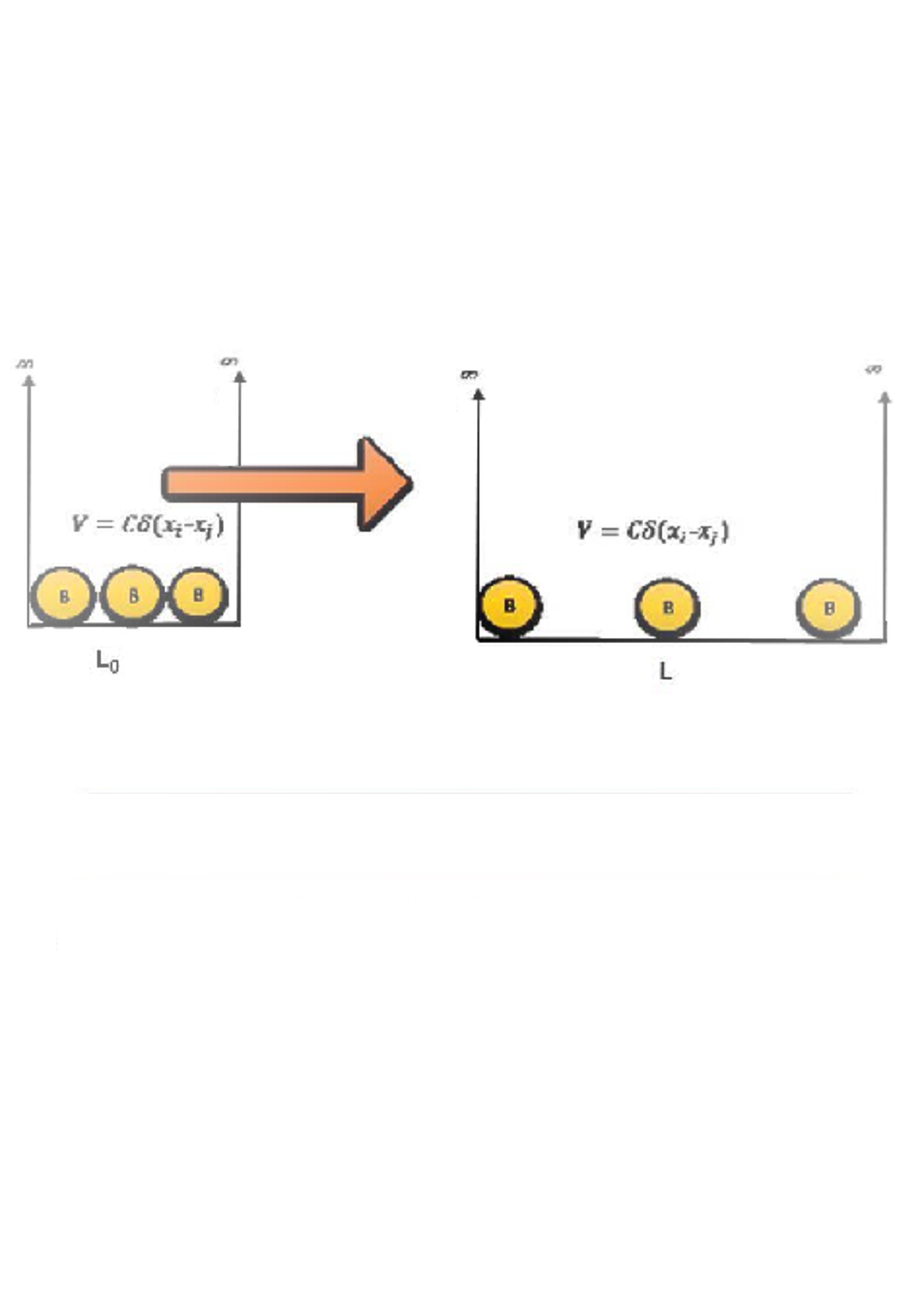}
\vspace{-64.5 mm}
\caption{A plot for the thought experiment of expansion of the gas from the box of Length $L_0$ to $ L$}
\end{figure}
\clearpage
The subsequent time evolution of the many-body state
is governed by
\begin{equation}
\psi(x_1,...,x_N,t)
=
e^{-iHt/\hbar}
\psi_0(x_1,...,x_N).
\end{equation}

This sudden change of the trap geometry constitutes a
quantum quench that initiates the expansion dynamics.

\section{Density Dynamics}

The many-body ground state and time evolution are
computed using a quantum Monte Carlo technique based
on the generalized Feynman–Kac path integral
representation.

In imaginary time the Schrödinger equation can be
written as

\begin{equation}
\frac{\partial \psi}{\partial t}
=
\left(
\frac{\Delta}{2}
-
V
\right)
\psi .
\end{equation}

The solution admits the stochastic representation

\begin{equation}
\psi(x,t)
=
E_x
\left[
e^{-\int_0^t V(X(s))ds}
f(X(t))
\right],
\end{equation}
here $X(t)$ denotes a diffusion process.
This representation allows the ground state and time-dependent wavefunctions to be obtained
numerically by sampling stochastic trajectories. To speed up the convergence one can formulate the generalized Feynman-Kac method we first rewrite the Hamiltonian as $H=H_0+V_p$,
where $ H_0=-\frac{\Delta}{2}+{\lambda}_T+\frac{{\Delta}{\phi}_T}{2{\phi}_T}$ and $V_p=V-({\lambda}_T+\frac{{\Delta}{\phi}_T}{2{\phi}_T})$. We choose the non-negative trial function to be a trial function associated with the symmetry of the problem i.e.,$\phi={\phi}_T$ and ${\lambda}_T$ is the trial energy of this reference function.
Then the new stochastic solution can be written in terms of the reference potential $V_p$ as follows:
\begin{equation}
\nu(x,t)=E_{x}[e^{-\int_{0}^{t}{V_p}(Y(s))ds}]
\end{equation}
The diffusion $ Y(t) $ solves the following stochastic differential equation:\\
$dY(t)=\frac{\nabla {\phi_T}(Y(t))}{{\phi_T}(Y(t))}+dX(t)$.\\
${V_p}(Y(s))$ is summed over all the time steps and $e^{-{V_p}(Y(s))}$ is summed over all the trajectories.

The many-body spatial density is then calculated as

\begin{equation}
\rho(x,t)
=
N_c
|\psi(x_1,...,x_N;t)|^2 .
\end{equation}

This approach provides an ab initio framework
for studying the nonequilibrium dynamics of
interacting quantum gases.

\subsection{Spatial Regions and Observables}

To quantify the redistribution of particles, we partition the system into two regions:

\begin{itemize}
\item Initial region: $[0,L_0]$
\item Expanded region: $[L_0,L]$
\end{itemize}

We define the average densities in these regions as
\begin{equation}
\rho_{L_0}(t) = \frac{1}{L_0} \int_0^{L_0} \rho(x,t)\,dx,
\end{equation}
\begin{equation}
\rho_{L}(t) = \frac{1}{L - L_0} \int_{L_0}^{L} \rho(x,t)\,dx.
\end{equation}

These quantities characterize the flow of particles between the two regions and serve as the basis for constructing the distance function used to analyze relaxation dynamics.
The stationary state considered here corresponds to a long-time dephased state rather than a thermal equilibrium state, consistent with the integrable nature of the system.
\subsection{Numerical Implementation}

We verify numerical accuracy by checking:

\begin{itemize}
\item stability under variation of the time step,
\item and saturation of observables at long times.
\end{itemize}

These checks are essential for reliably resolving subtle features such as the crossing of relaxation trajectories.
\section{Distance Function and Mpemba Criterion}

A central issue in identifying the Mpemba effect is the definition of an appropriate distance function that quantifies how far a system is from its long-time stationary state. Unlike equilibrium thermodynamics, where temperature or energy may serve as natural measures, in nonequilibrium quantum systems --- particularly integrable ones --- no unique choice of distance exists. The identification of a Mpemba effect is therefore inherently dependent on the observable used to characterize relaxation.

\subsection{Observable for Density Redistribution}

In the present setup, the dynamics is governed by a sudden expansion of the confining box, which induces a redistribution of particles from the initial region $[0,L_0]$ into the expanded region $[L_0,L]$. A natural observable for this process is the difference in spatially averaged densities between the two regions.

We define
\begin{equation}
D(t) = \left| \rho_{L_0}(t) - \rho_{L}(t) \right|,
\end{equation}
where $\rho_{L_0}(t)$ and $\rho_{L}(t)$ are the average densities in the initial and expanded regions, respectively.

This quantity provides a direct measure of the imbalance between the two regions and therefore captures the extent of density redistribution. At long times, due to dephasing, the system approaches a stationary state in which this imbalance becomes time-independent and typically small.

\subsection{Distance from the Stationary State}

To characterize relaxation, we define the distance from the stationary state as
\begin{equation}
\mathcal{D}(t) = \left| D(t) - D(t_{\mathrm{final}}) \right|,
\end{equation}
where $t_{\mathrm{final}}$ denotes a sufficiently large time at which the observable has effectively saturated.

This definition ensures that $\mathcal{D}(t) \to 0$ as $t \to t_{\mathrm{final}}$, providing a consistent measure of relaxation. Importantly, the same definition is used for all initial states, ensuring a meaningful comparison between different symmetry sectors.

\subsection{Initial States and Relaxation Trajectories}

We consider two distinct initial states prepared in the smaller box:

\begin{itemize}
\item the many-body ground state,
\item an excited state .
\end{itemize}

Following the quench, each state evolves under the same Hamiltonian, producing two relaxation trajectories:
\begin{equation}
\mathcal{D}_{\mathrm{gnd}}(t), \qquad \mathcal{D}_{\mathrm{exc}}(t).
\end{equation}

These trajectories quantify how each initial state approaches its respective stationary value in terms of density redistribution.

\subsection{Mpemba Criterion}

The Mpemba effect is identified through a reversal in the ordering of these distance functions. Specifically, a Mpemba-type effect occurs if

\begin{equation}
\mathcal{D}_{\mathrm{exc}}(0) > \mathcal{D}_{\mathrm{gnd}}(0),
\end{equation}
but at later times
\begin{equation}
\mathcal{D}_{\mathrm{exc}}(t) < \mathcal{D}_{\mathrm{gnd}}(t),
\end{equation}
for some $t > 0$.

To quantify this behavior more clearly, we define
\begin{equation}
\Delta(t) = \mathcal{D}_{\mathrm{gnd}}(t) - \mathcal{D}_{\mathrm{exc}}(t).
\end{equation}

A Mpemba effect is then characterized by:

\begin{itemize}
\item $\Delta(0) > 0$,
\item the existence of a crossing time $t_c$ such that $\Delta(t_c) = 0$,
\item $\Delta(t) < 0$ for $t > t_c$.
\end{itemize}

A robust Mpemba effect further requires that $\Delta(t)$ remains negative at later times without recrossing.

\subsection{Observable Dependence and Non-Universality}

It is important to emphasize that the Mpemba effect is not a universal feature of the dynamics but depends sensitively on the choice of observable. In the present case, the observable $D(t)$ captures density redistribution between spatial regions, and the observed crossing arises from differences in how the initial states populate the post-quench modes.

Different choices of observable --- for instance, local densities at specific points or global quantities such as total energy --- may not exhibit such a crossing. This highlights that the Mpemba effect should be understood as an observable-dependent phenomenon rather than a general law of relaxation.

\subsection{Physical Interpretation}

The origin of the Mpemba effect in this system can be traced to the interplay between initial state structure and the spectrum of post-quench modes. The ground state, characterized by strong spatial correlations and sharper confinement, appears to undergo a relatively rapid initial redistribution followed by slower relaxation. In contrast, excited states typically exhibit smoother relaxation due to their broader momentum distribution.

This difference in relaxation pathways leads to a crossing of the distance functions, even though both states evolve under the same Hamiltonian. The effect is thus rooted in the structure of the initial conditions rather than any dissipative mechanism.
\newpage
\section{Numerical Results and discussions}

In this section, we present the numerical results for the relaxation dynamics of the density observable. We consider different numerical resolutions and sampling regions to assess the robustness of the observed behavior.

\subsection{Density Observable: Ground and Excited States}



\begin{figure}[h]
\centering
\includegraphics[width=0.7\textwidth]{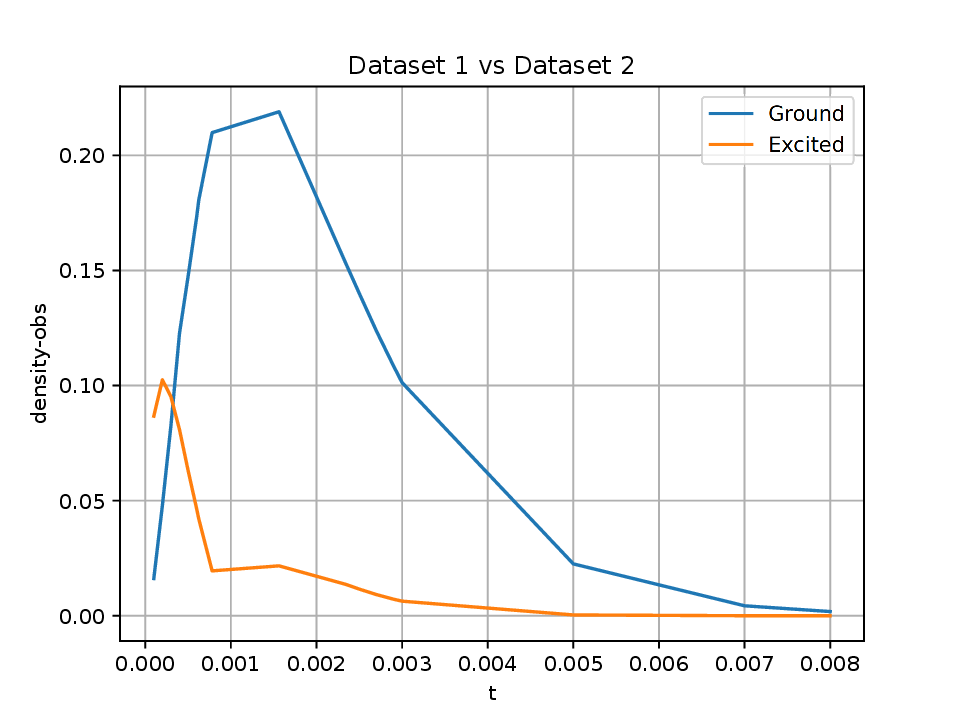}
\caption{Early-time evolution of the density observable showing a crossing and reversal in relaxation ordering for smaller step size.}
\end{figure}


\begin{figure}[h]
\centering
\includegraphics[width=0.7\textwidth]{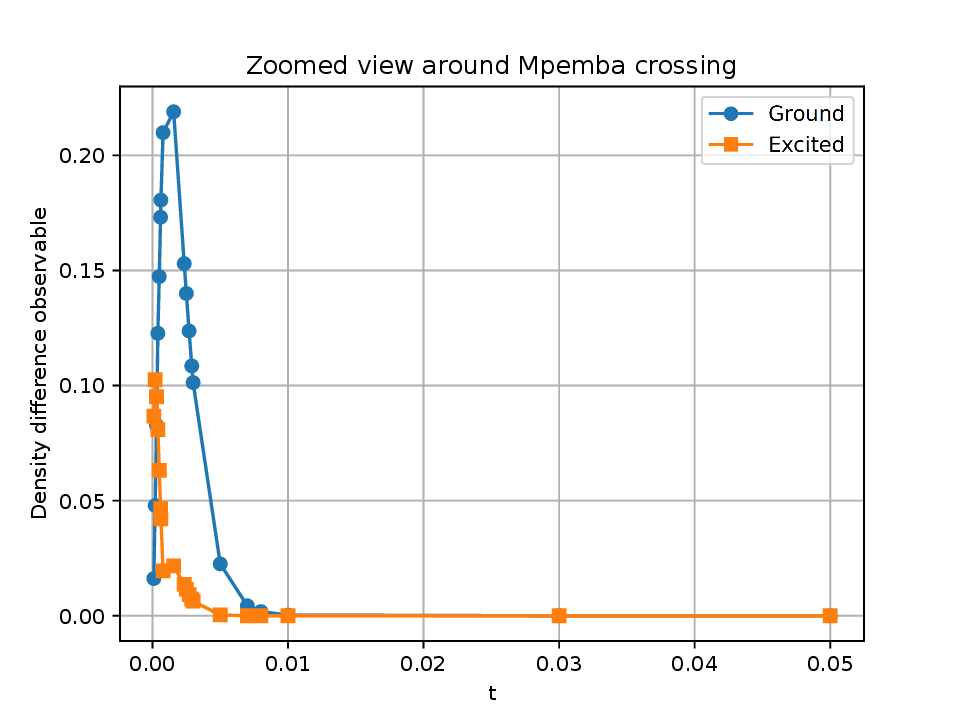}
\caption{Zoomed view of crossing region for a smaller time step showing reversal in relaxation ordering}
\end{figure}
\begin{figure}[h]
\centering
\includegraphics[width=0.7\textwidth]{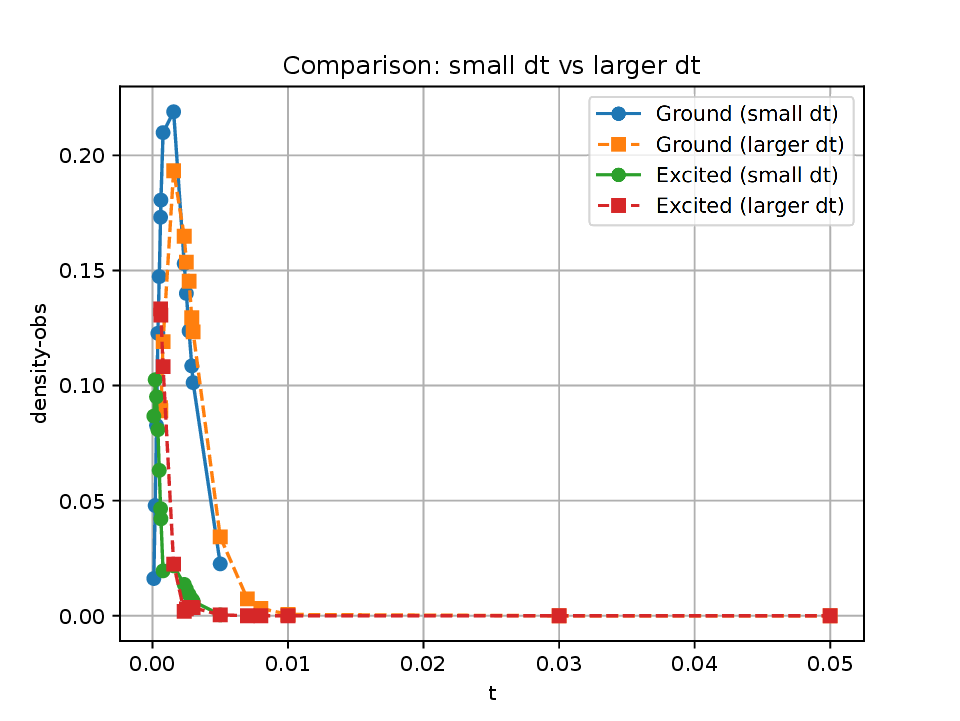}
\caption{Comparison of consistency of the crossing behavior at different time steps.}
\end{figure}
\newpage

\begin{figure}[h]
\centering
\includegraphics[width=0.7\textwidth]{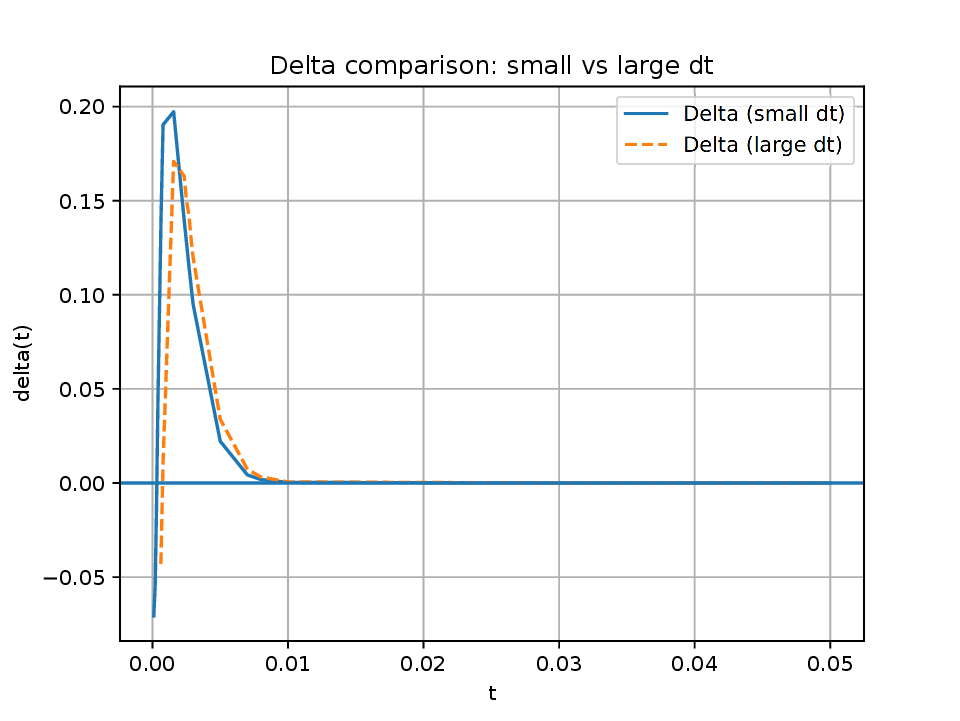}
\caption{Comparison of $\Delta(t)$ for two different time steps, showing consistency of the relaxation behavior.}
\end{figure}
\subsection{Description of Figures and Physical Significance}

We now describe the numerical results presented in Figs.~2--5 and discuss their physical implications in the context of Mpemba-type relaxation.

\paragraph{Figure 2: Early-time relaxation and onset of crossing.}
Figure~2 shows the time evolution of the density-based observable $D(t)$ for the ground and excited states at a finer temporal resolution in the early-time window. The two curves initially follow distinct relaxation trajectories, reflecting differences in the initial state structure. A crossing is observed within a short time interval, indicating a reversal in the ordering of relaxation. This constitutes the first evidence of Mpemba-type behavior in the present system. The early-time dynamics is particularly sensitive to the spatial redistribution induced by the sudden expansion.

\paragraph{Figure 3: Zoomed view of the crossing region.}
Figure~3 provides a magnified view of the crossing region identified in Fig.~2. The zoomed plot confirms that the crossing is well-resolved and not an artifact of insufficient resolution. The smooth behavior of both curves near the crossing point indicates that the reversal in relaxation ordering is a genuine dynamical feature. This figure establishes the existence of a well-defined crossing time $t_c$.

\paragraph{Figure 4: Robustness with respect to time step.}
Figure~4 compares the relaxation dynamics obtained using two different time steps. Both the ground and excited state trajectories exhibit qualitatively identical behavior across the two resolutions. In particular, the location of the crossing and the overall decay profile remain largely unchanged. This demonstrates that the observed Mpemba-type effect is not a numerical artifact arising from discretization, but a stable feature of the dynamics.

\paragraph{Figure 5: Difference function $\Delta(t)$ and Mpemba criterion.}
Figure~5 shows the time evolution of the difference function $\Delta(t) = D_{\mathrm{gnd}}(t) - D_{\mathrm{exc}}(t)$ for two different time steps. The function $\Delta(t)$ is initially positive, consistent with the excited state being farther from the stationary state at $t=0$. It crosses zero at a well-defined time and remains negative thereafter, indicating that the excited state relaxes faster than the ground state beyond the crossing point. The agreement between the two time-step results further confirms the robustness of this behavior.

\paragraph{Overall interpretation.}
Taken together, these figures provide consistent evidence of a Mpemba-type effect in the present system. Figure~2 establishes the existence of a crossing, Fig.~3 verifies its resolution, Fig.~4 demonstrates numerical stability, and Fig.~5 confirms the Mpemba criterion through the behavior of $\Delta(t)$. The results highlight that the observed effect arises from differences in the relaxation pathways of the initial states rather than from any numerical or transient artifact.
The numerical results presented in Sec.~IV indicate the presence of a crossing in the relaxation trajectories of the density-based distance function for different initial states. In this section, we analyze the robustness and physical interpretation of this observation.

\subsection{Robustness with Respect to Time Step}

A key aspect of the present study is the comparison of results obtained using different numerical time steps. Figures~6 and 10 show that the qualitative features of the dynamics, including the existence of a crossing in the relaxation curves and the behavior of the difference function $\Delta(t)$, remain consistent when the time step is varied.

In particular, we observe that:

\begin{itemize}
\item The crossing time $t_c$ is only weakly affected by the change in time step.
\item The ordering reversal between the ground and excited states persists across different resolutions.
\item The function $\Delta(t)$ exhibits a single zero crossing and remains negative thereafter in both cases.
\end{itemize}

These observations indicate that the detected Mpemba-type behavior is not a numerical artifact arising from insufficient temporal resolution, but rather a stable feature of the dynamics within the present computational framework.

\subsection{Dependence on Sampling Region}

The appearance of the crossing is also found to depend on the choice of spatial regions used to define the observable. While the density difference between the initial and expanded regions provides a physically motivated measure of redistribution, its quantitative behavior can vary with the precise definition of the sampling domains.

This sensitivity reflects the fact that the Mpemba effect is inherently observable-dependent. The redistribution of density is not uniform across space, and different regions can exhibit different relaxation rates. As a result, the presence or absence of a crossing may depend on how the observable probes the system.

\subsection{Role of Interactions and Integrability}

In the present study, the interaction strength is kept finite but large, placing the system close to the Tonks--Girardeau regime. Importantly, the interaction is retained throughout the dynamics and is not switched off after the quench. The time evolution is computed using a path-integral Monte Carlo approach, which directly samples the many-body wavefunction.

The observed relaxation is therefore governed by coherent many-body dynamics rather than a mapping to non-interacting fermions. The emergence of a Mpemba-type effect in this setting suggests that such behavior may not be restricted to exactly solvable limits , but can also arise in strongly interacting systems where correlations play a significant role.

At the same time, the system remains effectively integrable, and relaxation proceeds through dephasing rather than dissipation. The resulting stationary state corresponds to a long-time averaged configuration rather than a thermal equilibrium in the conventional sense.

\subsection{Interpretation of the Crossing Behavior}

The crossing observed in the distance function can be understood as a consequence of the different initial structures of the ground and excited states. The ground state, being more localized in the initial box, undergoes a rapid initial redistribution followed by slower relaxation at later times. In contrast, the excited state, with a broader momentum distribution, exhibits a more gradual but sustained relaxation.

This difference in relaxation pathways leads to a transient reversal in the ordering of the distance functions. Importantly, this behavior does not contradict basic principles of relaxation but instead reflects the multi-mode structure of the dynamics.

\subsection{Relation to Classical Mpemba Effect}

It is worth emphasizing that the present results do not imply a universal Mpemba effect. Similar to classical systems, where the effect depends on the interplay of relaxation modes and initial conditions, the quantum Mpemba effect observed here arises only under specific conditions.

In particular, the choice of observable, the preparation of initial states, and the details of the dynamics all play a crucial role. This is consistent with recent studies in both classical and quantum systems, which highlight the non-universal and context-dependent nature of the Mpemba effect.

\clearpage
\newpage
\section{Conclusions and Outlooks}

In this work, we have investigated the relaxation dynamics of a strongly interacting one-dimensional Bose gas undergoing a sudden box expansion. Using a path-integral Monte Carlo approach, we computed the time evolution of the many-body density and analyzed the redistribution of particles between spatial regions.

By defining a distance function based on the difference in densities between the initial and expanded regions, we identified a crossing in the relaxation trajectories of the ground and excited states. This crossing is further supported by the behavior of the difference function $\Delta(t)$, which exhibits a single zero crossing followed by a persistent change in sign.

A central outcome of our study is that this Mpemba-type behavior is robust with respect to variations in numerical time step and is reproducible across different sampling configurations. At the same time, the effect is found to depend sensitively on the choice of observable and the spatial regions used in its definition.

Our results therefore support the view that the Mpemba effect is not a universal law of relaxation but rather an observable-dependent phenomenon that emerges under specific dynamical conditions. In the present case, the interplay between strong interactions, initial state structure, and spatial redistribution leads to distinct relaxation pathways that enable the effect.

While the system considered here is close to the Tonks--Girardeau regime, the dynamics is computed using a fully interacting many-body framework. This suggests that such behavior may persist beyond strictly solvable limits and indicates that Mpemba-type behavior can arise in strongly interacting systems within the present framework.

Further work is needed to systematically explore the dependence of the effect on interaction strength, system size, and choice of observable. In particular, identifying general criteria for the emergence of Mpemba-like behavior in quantum systems remains an open problem.

This work does not claim universality of the Mpemba effect, but rather demonstrates its emergence under specific observable and dynamical conditions.

\newpage
{\bf Acknowledgements}: The author(SD) would like to thank Alliance University for providing support for carrying out the research work.\\

{\bf Declaration of interests:} The author does not have no conflicts of interest to declare and there is no financial interest to report.\\

{\bf Data availability statement:} No data in this publication is to be made available under the study-participant privacy protection clause.\\


\newpage
\begin{thebibliography}{99}

\bibitem{1} E. B. Mpemba and D. G. Osborne, Phys. Educ.,{\bf 4},172(1969) 
\bibitem{2} Aristotle. Meterologica (Clarendon Press, 1923).
\bibitem{3} F.  Ares, P. Calabrese and S. Murciano, Nat Rev Phys 7, 451–460 (2025). https://doi.org/10.1038/s42254-025-00838-0	
\bibitem{4} P. Calabrese J. Stat. Mech.  034002(2026) DOI 10.1088/1742-5468/ae4bb6
\bibitem{5} A. Lasanta, F. V. Reyes, A. Prados, A. Santos, Phys. Rev. Lett. 119, 148001 (2017).
\bibitem{6} Z. Lu and O. Raz, Proc. Natl Acad. Sci. USA 114, 5083 (2017).
\bibitem{7} I. Klich, O. Raz, O. Hirschberg and M. Vucelja, Phys. Rev. X 9, 021060 (2019).
\bibitem{8} A. Kumar and J. Bechhoefer, Nature 584, 64 (2020).
\bibitem{9} J. Bechhoefer, A. Kumar and R. A. Chétrite,  Nat. Rev. Phys. 3, 534 (2021).
\bibitem{10} A. Santos, Phys. Rev. E 109, 044149 (2024).
\bibitem{11} T.V. Vu and H. Hayakawa, Phys. Rev. Lett. 134, 107101 (2025).
\bibitem{12} G. Teza, J. Bechhoefer, A. Lasanta, O. Raz and M. Vucelja, Preprint at https://arxiv.org/abs/2502.01758 (2025).
\bibitem{13} A. K. Chatterjee, S. Takada and H. Hayakawa, Phys. Rev. Lett. 131, 080402 (2023).
\bibitem{14} D. J. Strachan, A. Purkayastha and S. R. Clark,  Preprint at https://arxiv.org/abs/2402.05756 (2024).
\bibitem{15} A. Nava and R. Egger, Phys. Rev. Lett. 133, 136302 (2024).
\bibitem{16} F. Ares, S. Murciano and P. Calabrese,  Nat. Commun. 14, 2036 (2023).
\bibitem{17} L. K. Joshi et al., Phys. Rev. Lett. 133, 010402 (2024).
\bibitem{18} J. Zhang et al.,  Nat. Commun. 16, 301 (2025).	
\bibitem{19} S. A. Shapira at al., Phys. Rev. Lett. 133, 010403 (2024).
\bibitem{20} D. Qian, H. Wang and J. Wang, Preprint at https://arxiv.org/abs/2411.18417 (2025).
\bibitem{21} M. Rigol, V. Dunjko and M. Olshanii, Nature 452, 854 (2008).
\bibitem{22} F. Ares, S. Murciano, E. Vernier and P. Calabrese, SciPost Phys. 15, 089 (2023).
\bibitem{23} C. Rylands et al., Phys. Rev. Lett. 133, 010401 (2024).
\bibitem{24} P. Calabrese and J. Cardy, J. Stat. Mech. 2005, P04010 (2005).
\bibitem{25} S. Murciano, F. Ares, I. Klich and P. Calabrese,  J. Stat. Mech. 2024, 013103 (2024).
\bibitem{26} K. Chalas, F. Ares, C. Rylands and P. Calabrese,  J. Stat. Mech. 2024, 103101 (2024).
\bibitem{27} S. Yamashika, F. Ares, and P. Calabrese,  Phys. Rev. B 110, 085126 (2024).
\bibitem{28} S. Yamashika, S., P. Calabrese and F. Ares, Phys. Rev. A 111, 043304 (2025).
\bibitem{29} F. Caceffo, S. Murciano and V. Alba, J. Stat. Mech. 2024, 063103 (2024).
\bibitem{30} X. Turkeshi, P. Calabrese and A. D. Luca, Preprint at https://arxiv.org/abs/2405.14514 (2024).
\bibitem{31} S. Liu, H.-K. Zhang, S. Yin, S.-X Zhang,  Phys. Rev. Lett. 133, 140405 (2024).
\bibitem{32} K. Klobas, C. Rylands and B. Bertini,  Phys. Rev. B 111, L140304 (2025).
\bibitem{33} H. Yu, Z.-X Li and S.-X Zhang,  Preprint at https://arxiv.org/abs/2501.13459 (2025).
\bibitem{34} F. Ares, S.  Murciano, P. Calabrese and L. Piroli,  Preprint at https://arxiv.org/abs/2501.12459 (2025).
\bibitem{35} S. Liu, H.-X Zhang, S. Yin, S.,  S.-X Zhang and H. Yao,  Preprint at https://arxiv.org/abs/2408.07750 (2024).
\bibitem{36} A. Foligno, P. Calabrese and B. Bertini, PRX Quantum 6, 010324 (2025).
\bibitem{37} G. Kells, D. Meidan and A. Romito, SciPost Phys. 14, 031 (2023).
\bibitem{38} M. Fava, L. Piroli, T. Swann, D. Bernard and A. Nahum, Phys. Rev. X 13, 041045 (2023).
\bibitem{39} F. Ares, V. Vitale and S. Murciano, Rev. B 111, 104312 (2025).
\bibitem{40} M. Kac in Proceedings of the Second Berkley Symposium (Berkley Press, California, 1951)

\bibitem{41} M. Girardeau, J. Math. Phys.{\bf 1}, 516(1960) DOI: 10.1063/1.1703687(1960)

\bibitem{42} M. Rigol, A. Muramatsu, Phys. Rev. Lett., 94, 240403(2005). DOI: 10.1103/PhysRevLett.94.240403

\bibitem{43} E. H. Lieb, W.Liniger, Phys. Rev. {\bf 1963}, 130, 1605(1963). DOI: 10.1103/PhysRev.130.1605

\bibitem{44} R. P. Feynman, A. R. Hibbs,  Quantum Mechanics and Path Integrals. DOI: 10.1063/1.3048320
\bibitem{45} P. Claverie, M. Cafferel, J. Chem Phys. {\bf 88 }, 1088 (1988);{\bf 88}, 1100 (1988).DOI: 10.1063/1.454227
\bibitem{46} A. Korzeniowski, J. L. Fry, D.E. Orr, N. G. Fazleev, Phys. Rev. Lett., {\bf 69}, 893(1992). DOI: 10.1103/PhysRevLett.69.893
\bibitem{47} S. Datta, J. L. Fry, N. G. Fazleev, S. A. Alexander, R. L., Coldwell, Phys. Rev. A , {\bf 61}, 030502(2000). DOI: 10.1103/PhysRevA.61.030502
\bibitem{48} X Wang, J Su, J Wang - Phys. Rev. B, {\bf 113}, 045119(2026)
\bibitem{49} S.Liu, HK Zhang, S. Yin, SX Zhang, H. Yao  Sci Bull (Beijing). 2025 Dec 15;70(23):3991-3996. doi: 10.1016/j.scib.2025.10.017. Epub 2025 Oct 17. PMID: 41193302. 
\bibitem{50} A. Summer, M. Moroder, L. Bettmann, X. Turkeshi, I. Marvian, and J. Goold, arXiv:2507.16976[quant-ph](2025)  
\bibitem{51} M. Moroder, O. Culhane, K. Zawadzki and J. Goold, Phys.Rev. Lett.,{\bf 133},140404(2024)
\bibitem{52} J. Furtado and A. C. Santos, Annals of Physics,{\bf 480}, 170135(2025)
\bibitem{53} Dynamical Fermionization and Emergent Bethe Rapidity Structure in the Spatial Density of Cold quenched Lieb-Liniger gas, Sumita Datta, J. M. Rejcek, Rajasee Datta 
             and M. Olshanii, submitted to Zeitschrift fur Naturforschung - A(2026)
\end{thebibliography}
\end{document}